\documentclass[10pt,letterpaper,american]{article}
\usepackage[latin9]{inputenc}
\usepackage{color}
\usepackage{babel}
\usepackage{amsmath}
\usepackage{amssymb}
\usepackage{graphicx}
\usepackage[unicode=true,
 bookmarks=false,
 breaklinks=false,pdfborder={0 0 1},backref=false,colorlinks=true]
 {hyperref}
\hypersetup{
 citecolor=blue,urlcolor=blue}

\makeatletter

\pdfpageheight\paperheight
\pdfpagewidth\paperwidth

\usepackage{osameet3}
\usepackage{capt-of}
\usepackage{booktabs}

\usepackage{cite}
\usepackage[printonlyused]{acronym}
\acrodef{DM}{distribution matcher}
\acrodef{Hi-DM}{hierarchical DM}
\acrodef{PS}{probabilistic shaping}
\acrodef{PAS}{probabilistic amplitude shaping}
\acrodef{AWGN}{additive white Gaussian noise}
\acrodef{QAM}{quadrature amplitude modulation}
\acrodef{FEC}{forward error correction}
\acrodef{PAM}{pulse amplitude modulation}
\acrodef{LUT}{look up table}
\acrodef{MB}{Maxwell-Boltzmann}
\acrodef{ESS}{enumerative sphere shaping}
\acrodef{CCDM}{constant composition distribution matching}
\acrodef{IR}{information rate}
\acrodef{SNR}{signal to noise ratio}
\acrodef{PPM}{pulse position modulation}
\acrodef{invDM}{inverse DM}
\acrodef{TX}{transmitter}
\acrodef{RX}{receiver}
\acrodef{BER}{bit error rate}
\acrodef{SER}{symbol error rate}

\makeatother

\begin{document}
\title{Cost-Gain Analysis of Sequence Selection\\
 for Nonlinearity Mitigation}
\author{S. Civelli\textsuperscript{1,2,{*}} and M. Secondini\textsuperscript{2,3}}
\address{\textsuperscript{1}CNR-IEIIT, Via Caruso 16, 56122, Pisa, Italy,
\emph{stella.civelli@cnr.it}\\
\textsuperscript{2} TeCIP Institute, Scuola Superiore Sant'Anna,
Via G. Moruzzi 1, 56124, Pisa, Italy\\
\textsuperscript{3}PNTLab, CNIT, Via G. Moruzzi 1, 56124, Pisa, Italy}

\maketitle
\copyrightyear{2024}
\begin{abstract}
We propose a low-complexity sign-dependent metric for sequence selection
and study the nonlinear shaping gain achievable for a given computational
cost, establishing a benchmark for future research. Small gains are
obtained with feasible complexity. Higher gains are achievable in
principle, but with high complexity or a more sophisticated metric.\vspace*{-0.5ex}
\end{abstract}

\section{Introduction}

\vspace*{-1ex}
Probabilistic constellation shaping (PS) has become a standard for
high capacity coherent optical fiber communication systems, as it
allows to have arbitrary granularity in rate and decrease the signal-to-noise
ratio required to achieve a certain rate, simply using some symbols
more often than others \cite{bocherer2015bandwidth,buchali2015jlt,cho2019,roadmap24}.
The success of PS and the substantial benefits it provides have pushed
the desire for additional gains when the channel exhibits nonlinear
behavior \cite{geller2016shaping,fehenberger2016JLT}. Unfortunately,
many studies recently showed that the nonlinear shaping gain that
can be obtained with the conventional PS (i.e., assigning a probability
to the amplitudes) is negligible or limited to specific scenarios
(e.g., for short-distance single-span transmissions) \cite{civelli2022JLTBPS}.

Within this context, sequence selection has recently been proposed
to optimize PS in a high-dimensional signal space (on sequences of
symbols), enabling the transmission of sequences that are optimized
for specific goals or scenarios. Originally, sequence selection was
proposed as an information-theoretic approach to study the ultimate
capacity limits of optical fiber communication \cite{civelli2021ecoc,secondini2021ecoc,secondini2022jlt}.
Next, practical coded modulation schemes based on the sequence selection
concept have been proposed, including the list-encoding CCDM \cite{wu2022listCCDM}
and the bit scrambling (BS) sequence selection \cite{civelli2024sequenceJLT}.
All these methods generate $N_{t}$ different sequences of symbols
from the same stream of information bits, characterize them with a
metric, and select the best one for transmission.

Clearly, the effectiveness of sequence selection depends critically
on the metric used to select the best sequences. A good metric should
accurately predict the performance of a given sequence on the nonlinear
channel while maintaining low computational cost, as it should be
evaluated many times ($N_{t}$ for any transmitted sequence). Some
low-complexity metrics that depend only on the symbol energies have
been proposed, such as the energy dispersion index (EDI) \cite{wu2022listCCDM}
or the lowpass-filtered symbol-amplitude sequence (LSAS) \cite{askari2023lsas}.
However, metrics that are based only on symbol energies provide limited
nonlinear shaping gains, while more relevant gains require a metric
that accounts also for the signs of the transmitted symbols \cite{civelli2024sequenceJLT}.
A few attempts to design sign-dependent metrics have been made recently:
the nonlinear interference (NLI) estimated through a noiseless simulation
of the channel \cite{civelli2024sequenceJLT}, a perturbation-based
metric \cite{askari2024ecocposter}, and the dispersion-aware EDI
\cite{liu2024sequence}. While these metrics provide some gain, their
computational cost has not been studied in detail, and is expected
to be large.

This work, for the first time, provides an analysis of the nonlinear
shaping gain versus computational cost for a metric that estimates
the nonlinear interference (NLI) generated by a sequence of symbols
based on a low-complexity numerical algorithm recently proposed for
digital backpropagation (DBP) \cite{civelli2024coupled,civelli2024JLT_CBESSFM}.
Despite the metric being based on a low-complexity propagation algorithm,
the results show that the computational cost required to unlock the
full potential of sequence selection is still too high. Nevertheless,
they serve as a benchmark for future studies on sequence selection
and highlight the need for a less complex but equally accurate selection
metric. 

\section{Sequence Selection}

\vspace*{-1ex}
Sequence selection was originally proposed in\cite{secondini2022jlt}
as a method to evaluate the information rate (here referred to as
\emph{sequence selection bound}) achievable when using sequences of
symbols with an optimized distribution. The approach is used to establish
the ultimate potential of shaping on high-dimensional constellations
(sequences of symbols) and is not conceived for practical implementation.
Therefore, it employs the complex averaged NLI metric in \cite[Eq. (12)]{secondini2022jlt},
where the NLI for a given sequence is estimated by including the impact
of random adjacent sequences (inter-block NLI) averaged over several
($20$) realizations. Sequences with a metric below a certain threshold
are selected for transmission, and the rate loss due to discarding
the other sequences is subtracted from the information rate.

Later, the sequence selection concept was extended from a pure information-theoretic
application to a practical coded-modulation scheme \cite{civelli2024sequenceJLT}.
Here, in particular, we consider the sequence selection implementation
based on \emph{bit scrambling} (BS) \cite[Sec. IV C]{civelli2024sequenceJLT}
and the simplified NLI metric in \cite[Eq. (13)]{secondini2022jlt},
which does not account for inter-block NLI to reduce computational
complexity. Differently from the information-theoretic approach, BS
maps information on $N_{t}$ different sequences (prepending $\log_{2}N_{t}$
pilot bits to identify the adopted mapping rule), and selects the
one with minimum metric. In this case, there is no guaranteed threshold
on the metric value, and the information rate is decreased by the
use of pilot bits.

Despite the differences, both approaches estimate the NLI (though
on a different number of symbols) as $||\mathbf{x}-\mathbf{y}||^{2}$,
where $\mathbf{x}$ is the transmitted sequence and $\mathbf{y}$
the received sequence after a noiseless single-channel (emulated)
propagation. The numerical method used to emulate fiber propagation
determines the accuracy of the NLI estimation and the overall computational
cost, given that it should be applied to each tested sequence. For
example, the naive approach of using the conventional split step Fourier
method (SSFM) with many steps estimates NLI with high accuracy  but
has a very high cost. Here, to reduce the computational cost, we employ
the recently proposed coupled-band enhanced SSFM (CB-ESSFM) algorithm,
which combines an SSFM-like structure with $N_{\mathrm{st}}$ linear
and nonlinear steps, a simplified logarithmic perturbation model,
and subband processing with $N_{\mathrm{sb}}$ subbands \cite{civelli2024coupled,civelli2024JLT_CBESSFM}.

BS requires the evaluation of the NLI metric $||\mathbf{x}-\mathbf{y}||^{2}$
for $N_{t}$ different sequences of $N=n_{\mathrm{sxs}}n$ 4D samples,
$n$ being the number of symbols in the sequence and $n_{\mathrm{sxs}}$
the number of samples per symbol used to represent the propagating
signal. For each sequence, the evaluation requires $4N$ real multiplications
(RMs) for $||\cdot||^{2}$ plus the cost of the propagation algorithm
for $\mathbf{y}$ (detailed in \cite{civelli2024JLT_CBESSFM}), but
for the last two complex FFTs (one per polarization), which can be
avoided by evaluating the NLI directly in the frequency domain. The
cost of additions is considered negligible compared to that of RMs.
Overall, the number of RMs per each transmitted 2D symbol is
\begin{equation}
C=\frac{N_{t}n_{\mathrm{sxs}}}{2}\left(5N_{\mathrm{st}}\log_{2}\frac{N}{N_{\mathrm{sb}}}+2\log_{2}N+N_{\mathrm{st}}\frac{3N_{\mathrm{sb}}+1}{2}+\frac{20N_{\mathrm{sb}}N_{\mathrm{st}}+8}{N}+4\right)\quad\mathrm{RM/2D\,symb.}
\end{equation}
excluding the cost to generate the $N_{t}$ sequences $\mathbf{x}$,
which depends on the starting constellation (PS or uniform) and on
the specific implementation (e.g., the distribution matcher).

\section{Performance Analysis}

\vspace*{-1ex}
\begin{figure}
\centering
\begin{centering}
\includegraphics[width=1\columnwidth]{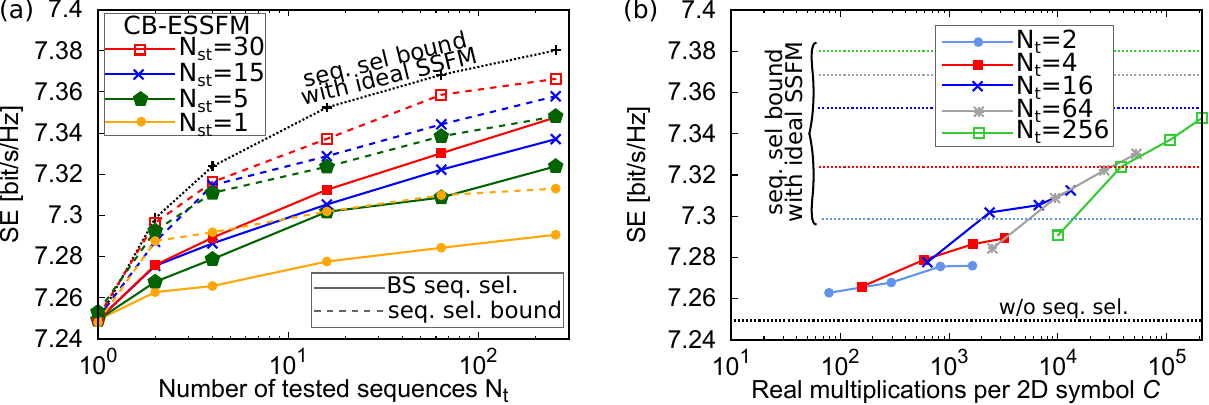}
\par\end{centering}
\caption{\label{fig:fig1}SE vs (a) number of tested sequences $N_{t}$ for
different numbers of CB-ESSFM steps $N_{\mathrm{st}}$; (b) total
computational cost for $N_{t}$ metric evaluations (with different
values of $N_{t}$).}

\vspace*{-2ex}
\end{figure}
The proposed CB-ESSFM metric is tested through simulations in the
same setup as \cite{civelli2024sequenceJLT}. The signal consists
of $5\times46.5$GBd dual polarization wavelength division multiplexing
channels with $50$GHz spacing, each modulated with a 64-quadrature
amplitude modulation with PS with rate $9.2$bits/4D and root-raised
cosine pulse (roll-off $0.05$). Sequence selection uses sequences
of $n=512$ 4D-symbols, starting from sphere shaping with block length
$256$. The metric is computed using $N_{\mathrm{sb}}=1$ and $n_{\text{sxs}}=1.125$.
The signal is sent into a channel made of $30\times100$km spans of
single mode fiber with erbium-doped fiber amplifiers (noise figure
$5$dB) which compensate for loss. At the receiver side, the central
channel, after demultiplexing, undergoes dispersion compensation,
matched filtering, symbol time sampling, and mean phase removal (in
this work, the impact of carrier phase recovery is not included, but
the results are expected not to change \cite[Fig. 13]{civelli2024sequenceJLT}).
The performance is given in terms of spectral efficiency (SE) in bit/s/Hz
at the approximately optimal launch power of $1$dBm.

Fig.~\ref{fig:fig1}(a) shows the SE obtained from the sequence selection
bound (dashed lines) and from the BS approach (solid lines), as a
function of the number of tested sequences $N_{t}$. The NLI is evaluated
using the CB-ESSFM with a different number of steps $N_{\mathrm{st}}$,
or using the ideal SSFM with many steps. The performance of all lines
improves when $N_{t}$ increases, as the selection becomes more accurate;
however it is expected to decrease for larger $N_{t}$ due to increased
information loss. As expected, the sequence selection bound is larger
than the SE obtained with BS with the same NLI metric. Furthermore,
the SE improves as $N_{\mathrm{st}}$ increases, remaining below that
obtained with ideal SSFM, which serves as benchmark. Interestingly,
when only a small number of test sequences $N_{t}$ is considered,
a lower-complexity metric can be employed (small $N_{\mathrm{st}}$)
to approach the ultimate gain achievable with an ideal metric; conversely,
for large $N_{t}$, higher gains can be achieved but only if using
a more complex metric (larger $N_{\mathrm{st}}$).

Fig.~\ref{fig:fig1}(b) reports the SE achieved with BS as a function
of the number of RMs required for the metric evaluation. The latter
is varied by changing either the number of tested sequences $N_{t}$
(different lines corresponding to different$N_{t}$ values) or the
accuracy of the selection metric (different points of the same line
corresponding to different number of CB-ESSFM steps $N_{\mathrm{st}}$).
The figure also reports the maximum SE achievable for each $N_{t}$,
corresponding to the sequence selection bound obtained with an ideal
(many steps) SSFM. The results show that small gains can be achieved
with feasible complexity, while higher gains are also achievable in
principle, but with higher complexity (or finding an accurate metric
with lower complexity). The results also indicate that, for a given
computational cost, the highest gain is obtained by selecting a proper
tradeoff between the number of tested sequences ($N_{t}$) and the
metric accuracy ($N_{\mathrm{st}}$). Overall, Fig.~\ref{fig:fig1}(b)
establishes a benchmark for cost-gain trade-off of sequence selection,
better metrics being required to push the solid lines towards the
dashed lines.

\section{Conclusion}

\vspace*{-1ex}
This work presents, for the first time, an analysis of the trade-off
between computational cost and nonlinear shaping gain obtained through
sequence selection. The results show that, with the currently proposed
selection metric, a small gain can be achieved with a feasible complexity.
On the other hand, higher gains are achievable with very high complexity
or, in principle, by finding a better metric with same accuracy and
lower complexity. The work establishes a benchmark for future research
on sequence selection, emphasizing the importance of finding better
low-complexity selection metrics to unlock the full potential of sequence
selection.

\section*{Acknowledgment}

\vspace*{-1ex}
This work was partially supported by RESTART PE00000001.

\end{document}